# Accepted Manuscript

Structural and electronic properties of the μ-phase Fe-Mo compounds

J. Cieslak, J. Przewoznik, S.M. Dubiel

PII: S0925-8388(14)01308-5
DOI: http://dx.doi.org/10.1016/j.jallcom.2014.05.201
Reference: JALCOM 31390

To appear in:

Received Date: 13 December 2013
Revised Date: 8 May 2014
Accepted Date: 27 May 2014
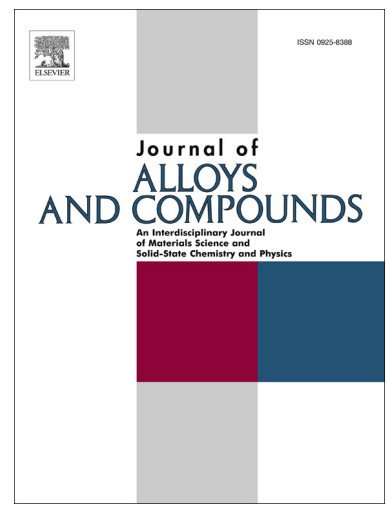

Please cite this article as: J. Cieslak, J. Przewoznik, S.M. Dubiel, Structural and electronic properties of the μ-phase Fe-Mo compounds, (2014), doi: http://dx.doi.org/10.1016/j.jallcom.2014.05.201

This is a PDF file of an unedited manuscript that has been accepted for publication. As a service to our customers we are providing this early version of the manuscript. The manuscript will undergo copyediting, typesetting, and review of the resulting proof before it is published in its final form. Please note that during the production process errors may be discovered which could affect the content, and all legal disclaimers that apply to the journal pertain.

# Structural and electronic properties of the μ-phase Fe-Mo compounds

J. Cieslak, J. Przewoznik and S.M. Dubiel[*]

*AGH University of Science and Technology, Faculty of Physics and Applied Computer Science, al. Mickiewicza 30, 30-059 Krakow, Poland*

Structural (lattice parameters and sub lattice occupancies) and electronic (charge-density and electric field gradient) properties in a series of μ-$Fe_{100-x}Mo_x$ (37.5≤$x$≤44.5) compounds were studied experimentally (X-ray diffraction and Mössbauer spectroscopy) and theoretically (charge and spin self-consistent Korringa–Kohn–Rostoker Green's function method). The lattice parameters $a$ and $c$ showed a linear increase with $x$ while all five lattice sites were found to be populated by both alloying elements: A(1a) and B(6h) predominantly by Fe atoms whereas C(2c) and D(2c') by Mo atoms hardly depending on the composition. The population of Fe atoms on the site E(2c'') was ranging between ~50% at $x$=37.5 and ~20% at $x$=44.5. Fe-site charge-density (isomer shift) and the electric field gradient (quadrupole splitting, *QS*) were revealed to be characteristic of the lattice site and both of them were almost $x$-independent. The difference in the charge-density at Fe-atoms at the sites B (the highest value) and those at the sites D (the lowest value) was estimated as high as 0.18 electrons. The average charge-density increases linearly with $x$. The largest *QS*-values were those at the sites A and C, while the smallest ones at the site D. The average *QS*-value was 0.25 mm/s.

Key words:

A. Intermetallic

C. Electronic properties, preferential site ordering

D. Mössbauer spectroscopy, X-ray diffraction,

[*]Corresponding author: Stanislaw.Dubiel@fis.agh.edu.pl

## 1. Introduction

The μ-phase was reported to be known in over 40 cases [1]. It belongs to a family of Frank-Kasper (FK) phases [2,3]. The FK-phases, including among others β, χ, δ, λ (Laves), η, σ, ν, *M, P, R* phases, are complex intermetallic compounds whose characteristic features are high values (12-16) of coordination numbers (CN) and mixed occupancy of sub lattices by constituting elements and a lack of a definite stoichiometry [4]. The latter can be of advantage, on one hand, as it enables smooth changing of physical properties by changing chemical composition of the compounds. On the other hand, however, it makes the FK-phases very complex and challenging as far as theoretical calculations as well as interpretation of experimental data (in particular those obtained with microscopic method) are concerned. They are also regarded as electron compounds [5,6]. The FK-phases have been of both technological and scientific interests. The former stems from the fact that precipitation of these phases in technologically important materials, e. g. σ in high-Cr steels or μ in Co, Fe and Ni based superalloys, drastically deteriorates their useful properties like corrosion resistance, creep strength, impact toughness or tensile ductility e. g. in [7-10]. In other words, the precipitation of the FK-phases is highly undesired from the industrial view-point, hence their formation should be avoided or, at least, controlled. On the contrary, for solid state and materials scientists the complexity of the FK-phases and a resulted diversity of their physical properties are highly challenging. The μ-phase (Pearson symbol: hR13, space group: $C_{3d}^5 - R\bar{3}m$, archetype: $Fe_7W_6$) has a rhombohedra structure and hosts 13 atoms distributed over 5 lattice sites with CN ranging from 12 to 16. Its unit cell can be also presented in hexagonal setting (39 atoms), where it is characterized by *c/a*≈5.3-5.5 [11]. Here reported are results obtained for two series of seven μ-$Fe_{100-x}Mo_x$ compounds, 37.5 ≤ *x* ≤ 44.5 i.e. the whole range of the μ-phase occurrence in the Fe-Mo alloy system was covered [12]. The results were obtained using experimental viz. X-ray diffraction (XRD) and Mössbauer spectroscopy (MS) techniques as well as theoretical viz. electronic structure calculations using the Korringa-Kohn-Rostoker (KKR) method. The XRD gave information on structural properties (lattice constants and sub lattice occupancies), while the MS measurements in combination with the KKR calculations yielded information on the Fe-site charge-density (isomer shift) and the electric field gradient (quadrupole splitting). The isomer shifts obtained for μ are compared with those determined for another two FK-phases present in the Fe-Mo alloy system viz. for σ and for *R*.

## 2. Experimental

### 2.1. Samples

Three series *S1*, *S2* and *S3* of $Fe_{100-x}Mo_x$ alloys with a nominal composition *x*=37.5, 38.0, 39.5, 41.0, 42.5, 44.0 and 44.5 covering the whole range of the μ-phase occurrence in the Fe-Mo alloy system [12] were prepared by pressing a 3g mixture of elemental iron (99.9%) and molybdenum (99.95%) in form of powder. Such prepared tablets were next heat treated as shown in Table 1.

Table 1

The following verification by means of XRD and MS gave evidence that none of the three series has been 100% transformed into µ (in *S1* and *S2* the transformation degree was ~90%, while in *S3* only ~10%). Consequently, all samples were re-powdered by attrition, re-pressed as before and they underwent a similar heat treatment as in the first step. The phase-verification of the samples after the second heat treatment gave clear evidence that those annealed at 1000 and 1200°C had been almost completely (≥97.5%) transformed into µ, while those annealed at 700°C still contained a significant fraction of the virgin α-phase. In these circumstances, the series *S3* has been excluded from a further study i.e. the samples indicated as *S1* and S2 after the second heat treatment were used in our measurements described below. The loss of mass after the applied heat treatment was negligible, consequently the nominal composition is used throughout the paper.

**2.2. XRD measurements**

X-ray diffraction (XRD) patterns were collected with an Empyrean PANalytical diffractometer using a diffracted beam graphite monochromator and an X'Celerator linear detector (Cu $K_\alpha$ radiation). The patterns were recorded at room temperature in the angular range of 10-120° (2θ) with steps of 0.0167° and were analyzed using the profile fitting program *FullProf* [13] based on the Rietveld method. The background intensity was refined with a polynomial and the peak shape was approximated with a pseudo-Voigt function. As an example, the Rietveld refined XRD pattern for the µ-$Fe_{59}Mo_{41}$ alloy is shown in Fig.1.

Fig. 1

In all samples of the series *S1* and in 3 samples of the series *S2* (x ≤ 39.5) trace amounts (< 2.5 wt %) of $Fe_2MoO_4$ were found. A Similar amount of $Fe_2Mo_3O_8$ was revealed in the other 4 samples of S2 (x > 39.5). Taking into account that the samples of both series were prepared under different protective atmospheres (argon for *S1* and vacuum for *S2*), and that the amount of oxides was similar in both cases, one can assume that the residual oxide phase was due to the oxygen that had been trapped by pressing the powders and/or the one existing in form of oxides on surfaces of the ingredients used for the fabrication of the samples. Whatever the source of the foreign oxide phases found in the final samples, their existence, due to a small amount, was neglected in the analysis of the measurements and interpretation of the data reported in this paper.

The analysis of XRD-patterns yielded values of the lattice parameters, *a* and *c* (Fig. 2a-c), atomic positions (Fig. 2d, Table 2), the average radius of the nearest-neighbor shell, *<d>* (Fig. 2e-f, Table 2), as well as occupancies of 5 sub lattices by Fe atoms present in the unit cell of µ (Fig. 3).

Table 2

Fig. 2

### 2.3. Mössbauer spectroscopic measurements

For the MS measurements the samples were used in form of powder with a density of 10 mg Fe/cm$^2$. The spectra were recorded at RT in 512 channels using a standard spectrometer with a sinusoidal drive and a $^{57}$Co/Rh source for the 14.4 keV gamma rays. Initially, both *S1* and *S2* series of the spectra were recorded within the velocity range of ±8 mm/s in order to verify whether or not a magnetic component exists.

As it has turned out that this was not the case, the spectra were measured again but this time, as shown in Fig. 4, within the velocity range of ±2 mm/s. It is evident that the spectra have a shape of doublets that for $x < 44$ are slightly asymmetric. It is also obvious that one cannot uniquely analyze such spectra in terms of five sub spectra corresponding to the five sub lattices present in the unit cell of µ. However, there are several possible ways of their analysis viz. either in terms of (a) the isomer shift (*IS*) distribution, (b) the quadrupole splitting (*QS*) distribution or (c) electronic structure calculation-aided superposition of five sub spectra. In the (a) and (b) cases only the average values of *IS* and *QS,* respectively*,* can be obtained. The (a) approach resulted in a better statistical quality of the fits than the (b) one. By integration of the *IS*-distribution curves, the average values of the isomer shift, *<IS>*, were calculated. The analysis of the spectra in which the electronic structure (charge-density and the electric field gradient) at Fe-atoms occupying particular lattice sites could be obtained was based on the theoretical calculations. Since no difference between the corresponding spectra recorded on the two series of samples was found, we present in the following only results obtained for the *S1*-series.

### 3. Theoretical calculations

The charge and spin self-consistent Korringa–Kohn–Rostoker (KKR) Green's function method [14-16] was used to calculate the electronic structure of the Fe–Mo µ-phase. The crystal potential of muffin-tin (MT) form was constructed within the local density approximation (LDA) framework using the Barth–Hedin formula [17] for the exchange–correlation part. The group symmetry of the unit cell of the µ-phase ($R\bar{3}m$) was lowered to allow for various configurations of Fe/Mo atoms. The experimental values of lattice constants and atomic positions were applied in all computations. For fully converged crystal potentials electronic density of states (DOS), total, site-composed and *l*-decomposed DOS (with $l_{max}$=2 for Fe and Mo atoms) were derived. Fully converged results were obtained for ~120 special ***k***-point grids in the irreducible part of the Brillouin zone but they were also checked for convergence using a denser ***k***-mesh. Electronic DOSs were computed using the tetrahedron ***k***-space integration technique and ~700 small tetrahedra. The KKR calculations were carried out for 85 unit cells with different configurations of Fe and Mo atoms on the sub lattices, keeping experimental lattice constants in all cases as constraints. The configurations were chosen in such a way that each possible number of Fe atoms being the nearest-neighbors for a given lattice site, $NN_{Fe}$, had been taken into account. More details relevant to the calculations can be found elsewhere [18,19].

### 4. Results and discussion

### 4.1. Structural properties

Analysis of the measured XRD patterns enabled determination of the hexagonal lattice parameters *a* and *c,* the average radius of the nearest-neighbor shell, atomic positions as well as sub lattice occupancies (positional occupancy factors for the sites) of the Fe and Mo atoms. Generally, it should be noted that all of the mentioned quantities show a linear change with increasing Mo content, *x*, and systematic differences of the corresponding data for the two series.

Concerning the lattice parameters, as shown in Fig. 2a,b, both *a* and *c* exhibit a linear increase with *x*: in the case of *c* also the effect of the heat treatment is visible. Similarly, the linear increase can be observed for the lattice parameter converted into the rhombohedral unit cell (Fig. 2c). Interestingly, there was a very slight change in the angle alpha (the biggest differences do not exceed 0.02°) which indicates a uniform change in the dimensions of the unit cell i. e. without any distortion of its shape. The increase of both lattice parameters with *x* can be understood in terms of the atomic radii of the compound constituting elements: 1.26Å for Fe and 1.39Å for Mo.

Atomic positions in the unit cell, $(x_k, z_k)$, show systematic changes with the increase of *x*, differences between the series *S1* and *S2* (after the second heat treatment) can also be seen. To illustrate this behavior, a corresponding dependence for the sub lattice E over the whole range of *x* is shown in Figure 2d. The results for each sub lattice and a selected concentration of *x*=41 can be found in Table 2. As can be seen, the differences between the values are small, and even in extreme cases, do not exceed ~0.3%. Average nearest-neighbor distances, $<d>_k$, *(k=S1,S2)*, are shown as well (Fig.2e). In this case the extreme difference reaches ~1%, but this effect may be due to the observed increase of the lattice parameters with *x*. The same data normalized to the value of the lattice constant of the rhombohedral unit cell, $a_R$, (Fig. 3f) show a similar tendency of changes, but again, the differences do not exceed ~0.3%.

Concerning the occupancies of the sub lattices, they all are populated by both kinds of atoms, yet their distribution, as shown in Fig. 3, is not uniform: Fe atoms mostly occupy the A and B sites while the C and D sites are predominantly populated by Mo atoms. The occupancies of these sites hardly depend on the alloy composition. Concerning the site E, for *x*=37.5 both elements have about the same share but an increase of the Mo concentration results in a linear decrease of the population of Fe atoms down to ~18% for *x*=44.5. An effect of the heat treatment is visible, especially for the sites A and E, yet it goes in the opposite direction: the population of Fe atoms on the site A in the samples annealed at 1000°C is lower and that on the site E higher than the corresponding figures found for the samples annealed at 1200°C. This effect of the heat treatment may likely be responsible for a difference in the lattice constant, especially $a_C$ and $a_R$, as evidenced in Fig. 2a, and 2c, respectively.

Fig. 3

Fig. 4

### 4.2. Electronic properties

Let us first discuss the calculated charge-densities, $\rho_e$ presented in Fig. 5 for each of the five sub lattices as a function of the number of Fe atoms in the first neighbor-shell, $NN_{Fe}$. Circles represent the $\rho_e$-values obtained for 85 different atomic configurations taken into account in the calculations and the solid lines stand for the best linear fits to the data. In the case of A, C, D and E sub lattices the data constitute one "cloud" and the best-fit lines shows decreasing tendency with $NN_{Fe}$. Its rate, however, weakens when going from A to E.

Fig. 5

For the site B one can observe two separate "clouds" of data. On the basis of the XRD measurements it is known that the occupancy of this sub lattice by Fe atoms is not 100%. Consequently, calculations were done for five Fe-atoms and one Mo-atom on that site. On the other hand, this lattice site can accommodate 6 atoms of the unit cell that are arranged as two parallel triangles each hosting 3 atoms. If a given triangle is fully occupied by Fe atoms then the $\rho_e$-values constitute the upper cloud (open green circles). If, however, one of the site in the triangle is occupied by Mo atom, the remaining two Fe atoms have different charge-densities: one of them almost similar to those form the upper "cloud" (shown as full black circles), and the other one much smaller – they form the lower "cloud" (shown with open red circles).

Based on these calculations and the lattice sites occupancies shown in Fig. 3, the average *IS*-values, *<IS>*, for each site and composition were calculated. They are displayed as solid lines in Fig.6 (left panel) from which it follows that the isomer shift is a characteristic feature of a given lattice site but hardly concentration dependent.

For the sites C and D the isomer shift is positive indicating that the charge-density at these sites is lower than that in a pure $\alpha$-Fe, while for the other three sites it is negative i.e. the charge-density at Fe-atoms occupying A, B and E sites is higher than that in $\alpha$-Fe. In other words, the highest charge-density was found on the B-sites (~0.13$e$ higher than in $\alpha$-Fe using a scaling constant from Ref. 24), and the lowest one on the sites D (~0.05$e$ lower than in $\alpha$-Fe using a scaling constant from Ref. 24). It should be noted that the difference in the charge-density between the two sites viz. ~0.18$e$ is extremely high as far as metallic systems are concerned. This, in our opinion, clearly shows why the FK-phases are termed as electronic compounds.

The present calculations also enabled determination of the *QS*-values for each sub lattice versus the Mo content, $x$. The results obtained using an extended point charge model outlined in detail elsewhere [18] are displayed in Fig. 7 as solid lines. It is evident that also the *QS*-values are characteristic of a given lattice site and they exhibit a weak tendency for an increase versus $x$ for all sites except B. Concerning their values, hence those of the underlying electric field gradient, they lie within the range of 0.05 mm/s for the site D and 0.4-0.45 mm/s for the sites A and C.

Fig. 6

Based on the calculated *IS*- and *QS*-values and taking into account the sites occupancies as determined from the XRD measurements (Fig. 3), the Mössbauer spectra were successfully analyzed in terms of five sub spectra in form of doublets whose line widths, Γ, were treated as free parameters. In the fitting procedure the relative intensities of the sub spectra were constant and equal to the Fe-site occupancies determined from the XRD patterns. The *IS*- and *QS*-values were treated as semi-free parameters, i.e. allowing no more than 10% deviation from their calculated values. With this procedure very good fits were obtained with Γ-values ranging between 0.30-0.32 mm/s.

The values of the *IS* and *QS* for each composition and lattice site obtained from the Mössbauer spectra, using fitting procedure described above are presented in Figs. 6a and 7, respectively, using various symbols and colors.

Based on the theoretical data presented in Fig. 6a as solid lines, an average value of the isomer shift, *<IS>*, was computed and displayed in Fig. 6b as a solid line, as well. Corresponding experimental values obtained by means of a model–independent method (a) mentioned in paragraph 2.3 are presented as squares. A good agreement between the two sets of the data can be seen. Furthermore, it is evident that *<IS>(x)* decreases linearly with *x* i.e. the Fe-site charge-density increases with the increase of Mo atoms concentration. For the sake of comparison, we have added in Fig. 6b the *<IS>*-values found previously for σ-FeMo compounds [20] as well as for *R*-FeMo compounds [21].

Finally, a similar comparison has been done for *QS*-values – see Fig. 7 (low panel). Theoretically obtained *QS*-values are presented as lines, while experimental ones as open symbols. The average *QS*-values, *<QS>*, are displayed in the high panel of Fig. 7: the theoretical ones as a solid line, while those obtained from the spectra using method (b) are indicated as full circles. The latter show a weak concentration dependence, contrary to the former that has a value of *<QS>*≈0.25 mm/s and it hardly depends on *x*. The *<QS>*-values also compare pretty well with the one of 0.27 mm/s determined for the μ-$Fe_{60}Mo_{40}$ assuming the spectrum consists of two single lines [22]. For the μ-$Fe_{50}Nb_{50}$ compound *<QS>*=0.21 mm/s was reported [23].

The comparison of the data for three different FK-FeMo phases as displayed in Fig. 6b gives a clear evidence that the Fe-site charge-density in the μ-phase, which increases with *x*, is the highest one, and that in the σ-phase is the lowest one among the three FK-phases in the investigated Fe-Mo alloys. The difference in the isomer shift between σ and μ for *x*≈45 corresponds to ~0.065e as determined with the scaling constant from Ref. 24. The Fe-site charge density in the *R*-phase is only slightly lower than that in the μ-phase. It is also of interest to notice that for the μ-$Fe_{60}Mo_{40}$ compound a value of *<IS>*=-0.27 mm/s was reported [22]. As seen in Fig. 6 (right panel), it is off the trend found in the present study what can likely follow from a simplified way of the spectrum analysis used in [22]. The *<IS>*=-0.29 mm/s as found for the μ-$Fe_{50}Nb_{50}$ compound [23] is also off the presently found behavior assuming their linear extension to *x*=50 is valid – see Fig. 6 (right panel). However in this case the deviation is likely due to a different alloying element (Nb instead of Mo).

## 5. Conclusions

The paper reports results on structural and electronic properties of µ-phase $Fe_{100-x}Mo_x$ compounds in the whole range of its occurrence ($37.5 \leq x \leq 44.5$). The results were obtained using both experimental measurements (XRD and Mössbauer spectroscopy) and theoretical (charge and spin self-consistent Korringa–Kohn–Rostoker Green's function method) calculations. Concerning the structural properties, (1) lattice parameters *a* and *c* were revealed to linearly increase with the Mo concentration, a behavior that reflects a relationship between the atomic radii of the alloy constituting elements (1.26 Å for Fe, against 1.39 Å for Mo), (2) all five lattice sites are populated by both elements: A (1a) and B (6h) predominantly (>80%) by Fe atoms, while C(2c) and D(2c') predominantly (>90%) by Mo atoms. The occupancy of site E(2c'') by Fe atoms shows a strong dependence on the composition decreasing linearly from ~50% for $x$=37.5 to ~20% for $x$=44.5. By combining the KKR-based calculations and experimental methods, quantities pertinent to the electronic structure of the studied alloys viz. Fe-site charge-density (isomer shift) and electric field gradient (quadrupole splitting) were determined. Both of them were revealed to be characteristic of a given lattice site. The highest charge-density was found at Fe atoms occupying the sites B and A and the lowest one at those present on the sites D. The estimated difference between the extreme cases corresponds to ~0.18*e*, a figure that justifies the name of electronic compounds as often the FK-phases are termed. Among the three different FK-phases that can be formed in the Fe-Mo alloy system, the µ-phase shows the highest charge-densities at Fe-sites followed by those of the *R*-phase and the σ-phase. The discontinuity in the isomer shifts observed at $x \approx 37$ (small) reflects the effect of a difference crystal structure between *R* and µ, while that at $x \approx 45$ (large) can be ascribed to the corresponding difference between µ and σ. The values of the quadrupole-splitting range between ~0.4-0.45 mm/s for the sites A and C and ~0.05 mm/s for the site D. The average value being equal to ~0.25 mm/s. The results presented in this paper can be used as an adequate basis for testing and verifying different theoretical models pertinent to such complex phases as µ, R and σ. In particular, it would be of interest to see whether or not they are able to predict the discontinuities in the charge-density observed in the present study at $x \approx 37$ and $x \approx 45$.

## Acknowledgements

The Ministry of Science and Higher Education, Warszawa, is acknowledged for a financial support.

**Figure captions**

Fig. 1

Rietveld refinement of the XRD pattern for the µ-Fe$_{59}$Mo$_{41}$ alloy measured at RT. The set of vertical bars below the pattern represents the Bragg peak positions corresponding to the refined µ-phase structure (space group $C_{3d}^5 - R\overline{3}m$, upper row) and Fe$_2$MoO$_4$ phase (space group $Fd3m$, lower row). The bottom blue line represents the difference between experimental and calculated patterns.

Fig. 2

(Left panel) The variation of the refined hexagonal ($a_H$, $c_H$) as well as rhombohedral ($a_R$) unit cell parameters. (Right panel) Examples of crystallographic position, $x_E$, average *NN*-distance, $<d_E>$, and the one normalized to the rhombohedral unit cell, $<d_E>/a_R$, for the E-site. The quantities are presented as a function of the Mo content, *x*, for all µ-Fe$_{100-x}$Mo$_x$ samples of the series *S1* (blue circles) and *S2* (red diamonds). Solid lines are the best linear fits to the data.

Fig. 3

Probability of finding Fe atom on a given site, *P(Fe)*, versus the Mo content, *x*, as obtained from the Rietveld refinements of XRD-patterns recorded on the *S1* and *S2* series after the second heat treatment. Open symbols and solid lines (linear fits to the data) represent series *S1* while full symbols and dotted line stand for series *S2* (in the latter case only the data for the sites A and E are shown). Error bars are smaller than the size of symbols.

Fig. 4

RT Mössbauer spectra recorded on µ-Fe$_{100-x}$Mo$_x$ samples (series *S1* after the second heat treatment). The best fits in terms of five sub spectra (doublets) are presented as solid lines. The sub spectra are displayed for *x*=37.5.

Fig. 5

Fe–site charge–density, $\rho_e$, as calculated for the lattice sites A, B, C, D and E vs. the number of Fe atoms in the nearest neighbor–shell, $NN_{Fe}$. Symbols represent the $\rho_e$–values obtained for particular atomic configurations, while solid lines stay for the best linear fits to the data. Colors and solid lines for B are explained in the text.

Fig. 6

(a) The average isomer shift, $<IS_i>$, for the five lattice sites (*i*=A, B, C, D, E) versus Mo content, *x*. Theoretically calculated values are represented by solid lines. Open symbols stay for the *IS*-values found from the analysis of the Mössbauer spectra as described in the text. (b) The average isomer shift, $<IS>$, obtained (i) by the integration of the *IS*-distribution curves (squares) and (ii) from theoretical calculations (solid line) versus Mo concentration, *x*, for µ. Experimental data for σ and *R* FK-phases are also displayed. The data for σ was taken from [20], and that for *R* from [21]. For comparison added are the $<IS>$-values found for µ-Fe$_{60}$Mo$_{40}$ [22] (asterix) and that for µ-Fe$_{50}$Nb$_{50}$ [23] (cross).

Fig. 7

(Low panel) The quadrupole splitting, $QS_i$, for all five sub lattices (i=A, B, C, D, E), versus Mo content, $x$. Theoretically calculated values are represented by solid lines while open symbols stay for the $QS$-values found from the analysis of the Mössbauer spectra as described in the text. (Upper panel) The theoretically calculated average QS-value is indicated by a solid line while the full symbols stand for the average values of $QS$ obtained by integrating the $QS$-distribution curves - method (b) of spectra analysis.

**Table captions**

Table 1

Samples series and their thermal histories. After annealing the samples were dropped into liquid nitrogen (LN) or thrown in vacuum on a massive brass plate kept at room temperature (BR).

Table 2

Structural parameters of the µ-phase. Wyckoff Symbols (WS), crystallographic positions, ($x_k$, $z_k$), and the average nearest-neighbor distances, $<d>$, were calculated in rhombohedral unit cell from the present measurements, and are presented for the µ-$Fe_{59}Mo_{41}$ samples for the series S1 and S2 (after the second heat treatment). Values for $W_6Fe_7$ taken from H. Arnfeit and A. Westgren, *Jernkontorets Ann.* 119 (1935) 185

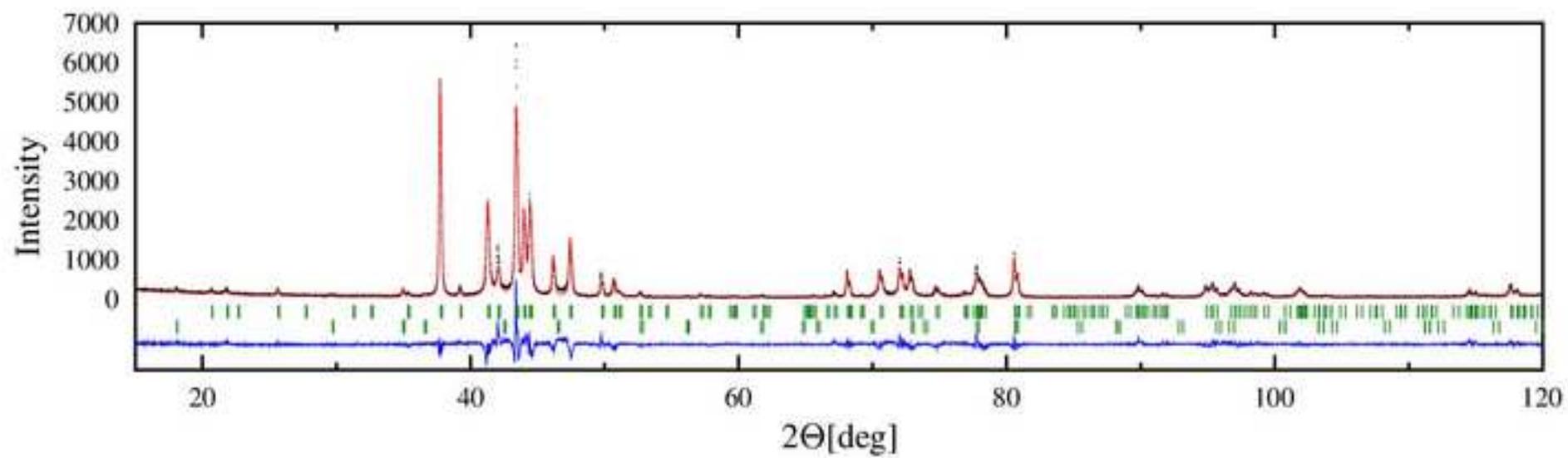

**Figure2**

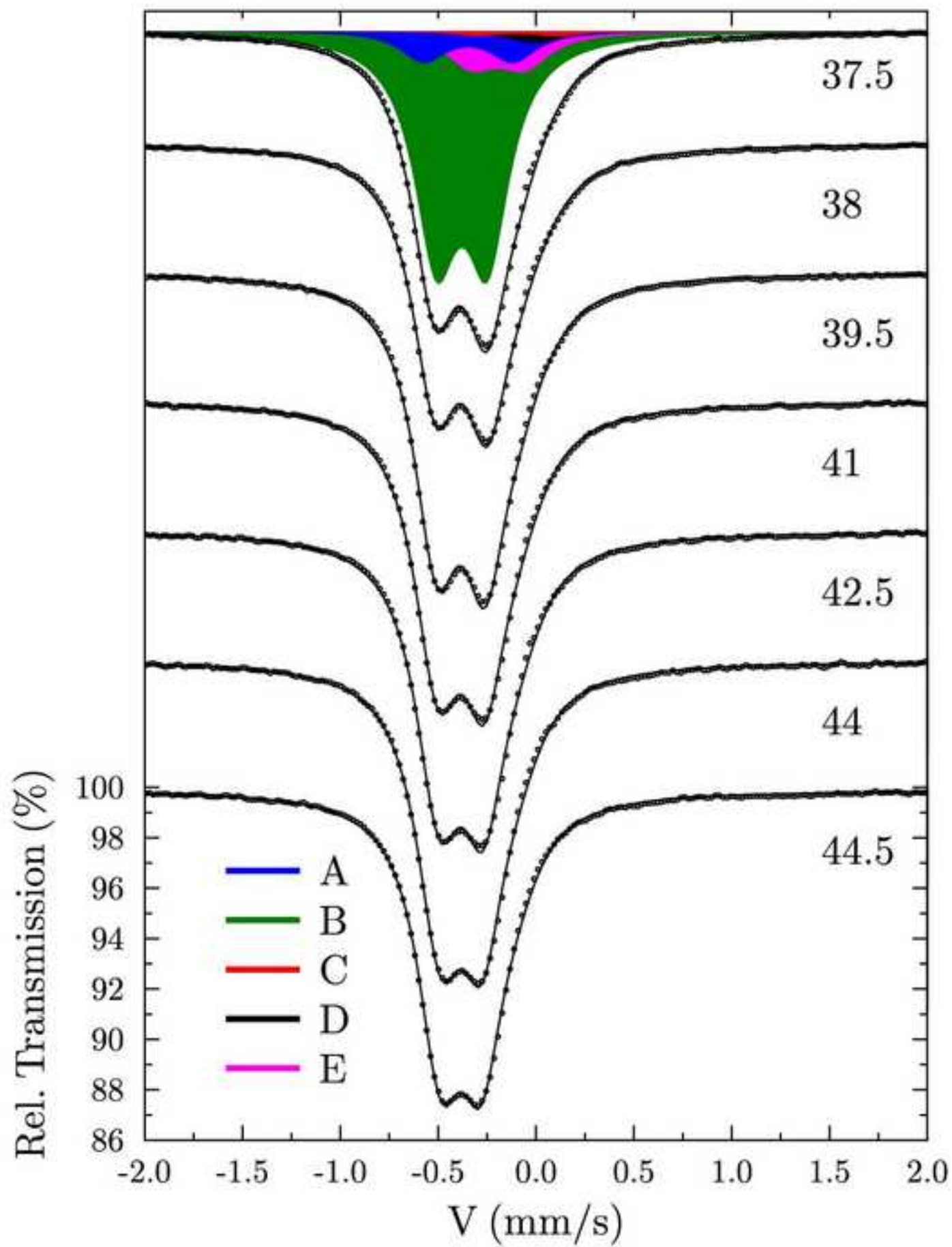

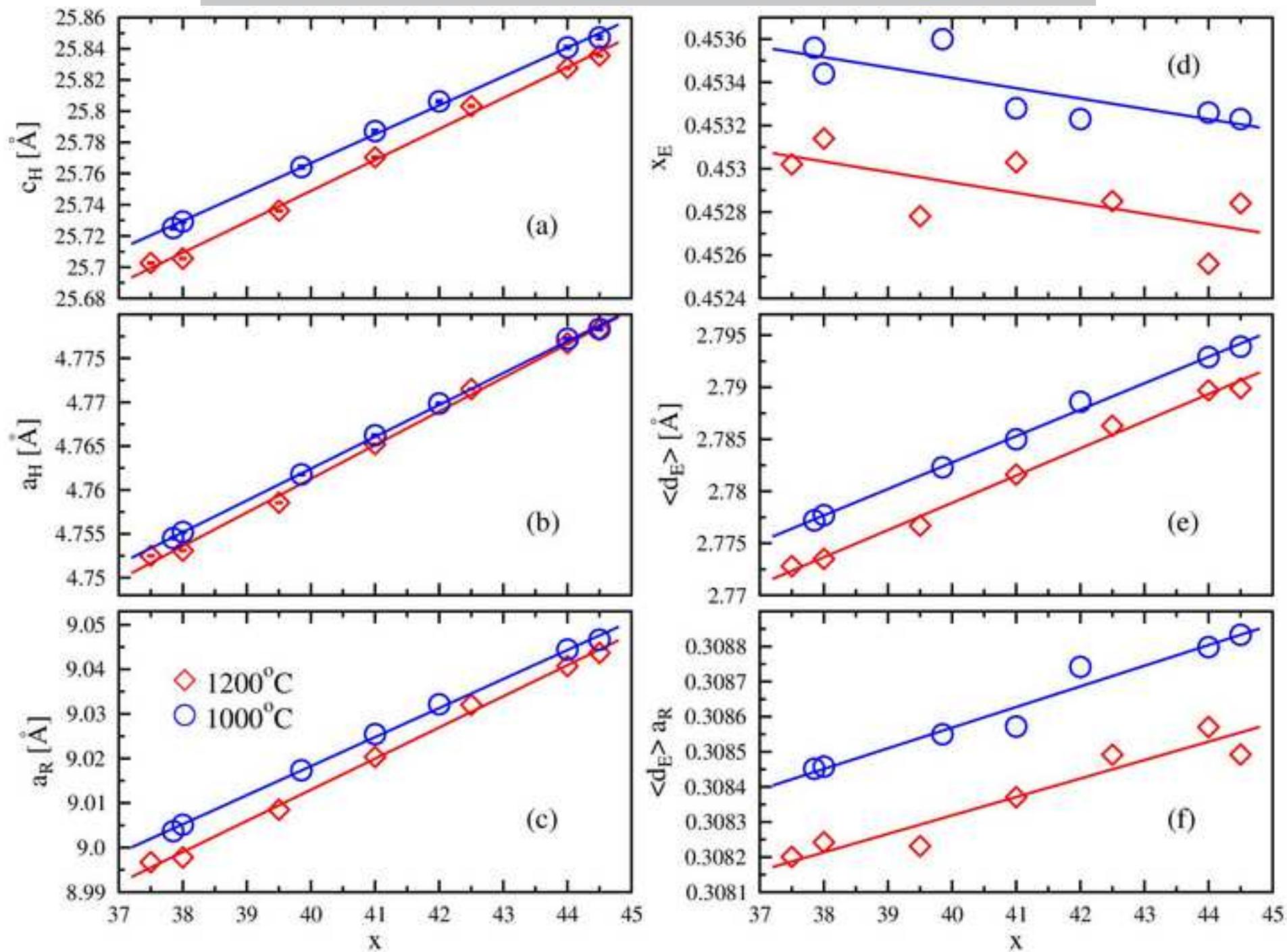

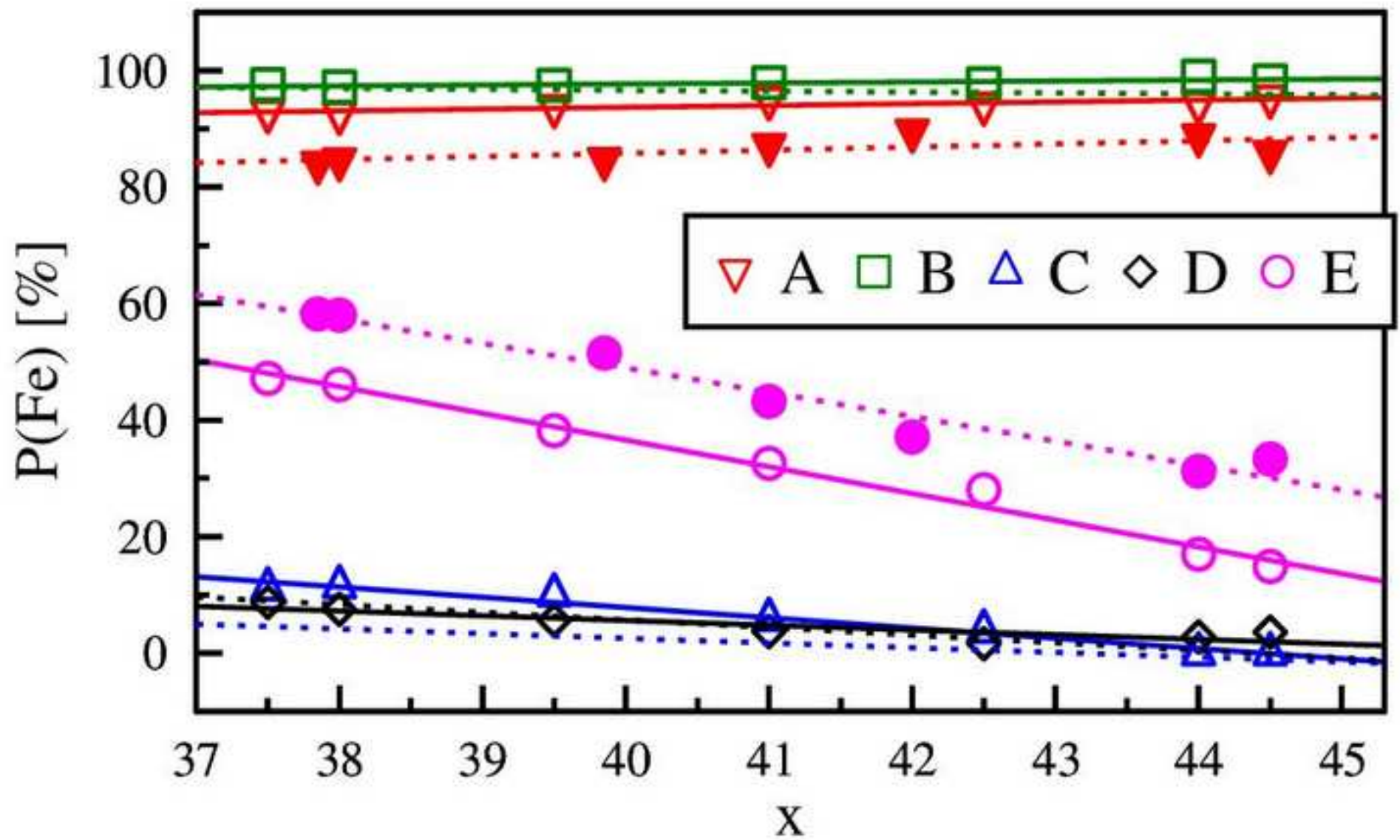

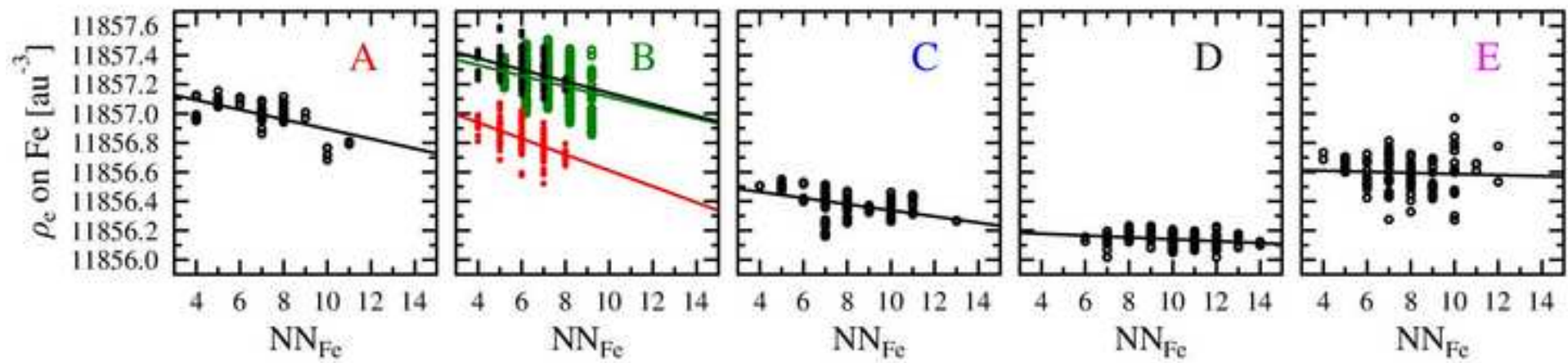

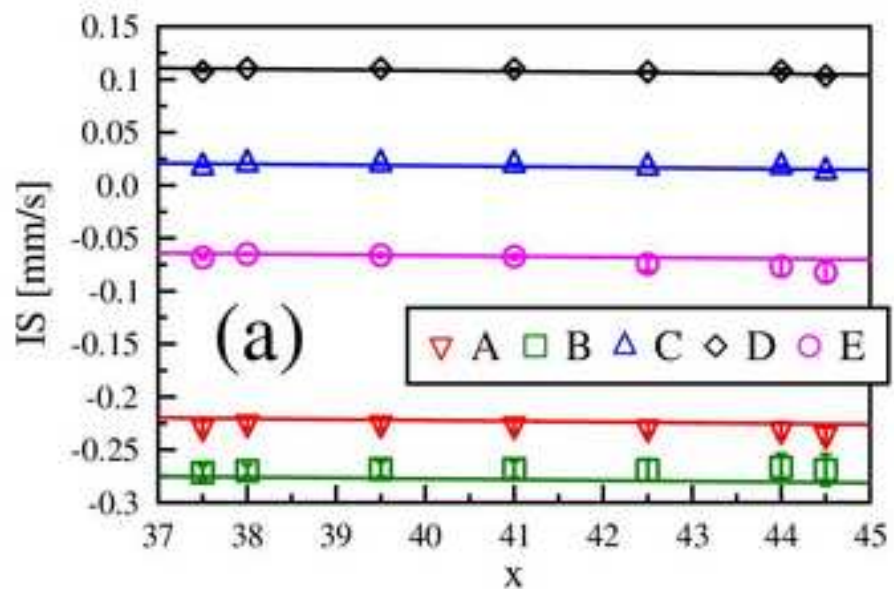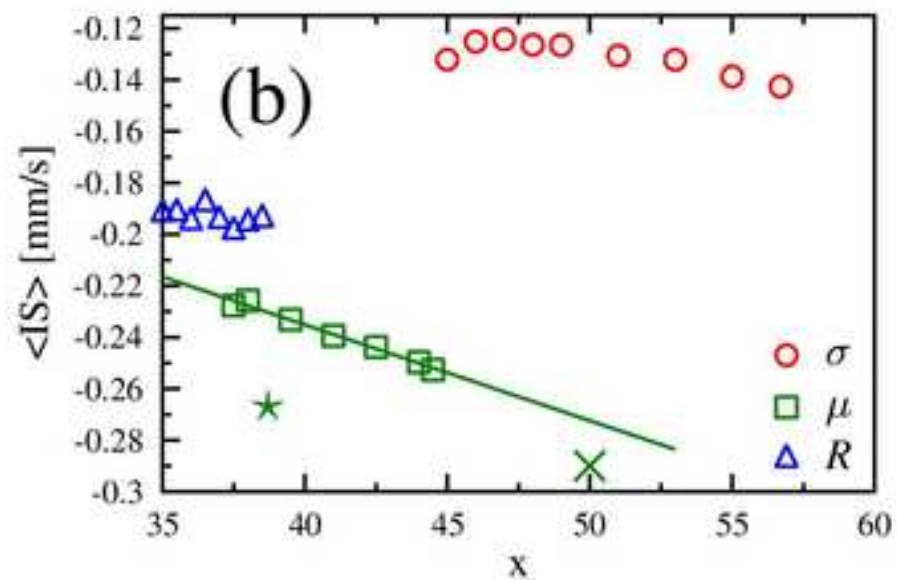

**Figure7**

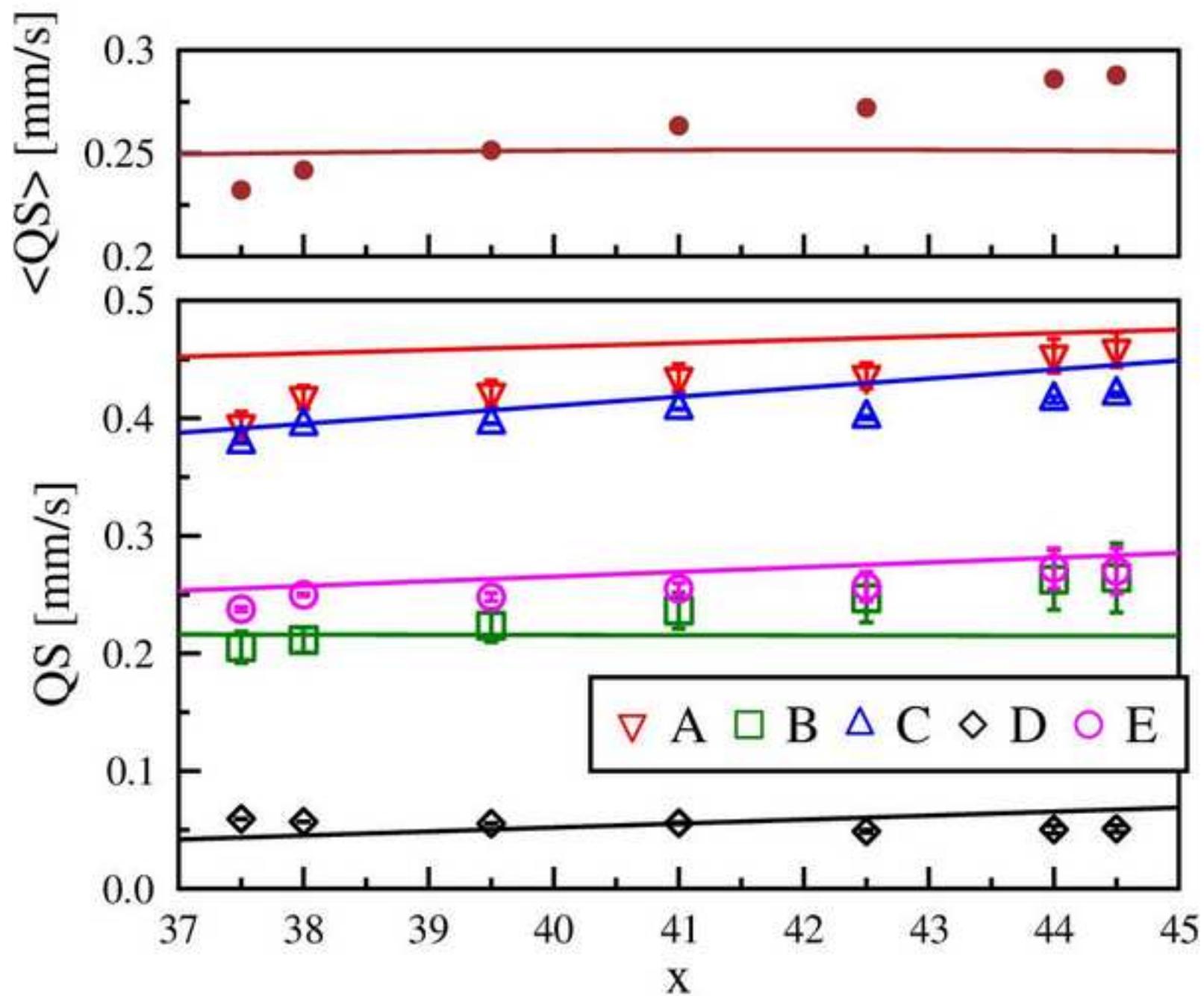

Table 1

| Series | $T_a$ [°C] | $t_a$ [h] | Atmosphere | Quenching |
|---|---|---|---|---|
| S1 | 1200 | 23 | Argon | LN |
| S2 | 1000 | 140 | Vacuum | BR |
| S3 | 700 | 140 | Vacuum | BR |

Table 2

| Site | WS | CN | NN-neighbors | | | | | $<d>_{S1}$ [Å] | $<d>_{S2}$ [Å] | Crystallographic positions | | | | | |
|---|---|---|---|---|---|---|---|---|---|---|---|---|---|---|---|
| | | | A | B | C | D | E | | | $x_{S1}$ | $z_{S1}$ | $x_{S2}$ | $z_{S2}$ | $x_{W6Fe7}$ | $z_{W6Fe7}$ |
| A | 1a | 12 | 0 | 6 | 0 | 6 | 0 | 2.598 | 2.594 | 0 | 0 | 0 | 0 | 0 | 0 |
| B | 6h | 12 | 1 | 4 | 2 | 3 | 2 | 2.574 | 2.576 | 0.0929 | 0.5843 | 0.0933 | 0.5848 | 0.09 | 0.59 |
| C | 2c | 16 | 0 | 6 | 3 | 0 | 6 | 2.832 | 2.835 | 0.1658 | | 0.1660 | | 0.167 | |
| D | 2c | 15 | 3 | 9 | 0 | 3 | 1 | 2.809 | 2.807 | 0.3481 | | 0.3471 | | 0.346 | |
| E | 2c | 14 | 0 | 6 | 6 | 1 | 1 | 2.782 | 2.785 | 0.4530 | | 0.4533 | | 0.448 | |

ACCEPTED MANUSCRIPT13

Highlights

• Experimental and theoretical study of µ-phase Fe-Mo compounds

• Sub lattice-resolved charge-densities and electric field gradients

• Extremely high values of Fe-site charge-densities